\documentclass{blois}

\usepackage{amsmath,amsfonts,amssymb}
\usepackage{mathtools}

\def\PLB{{\em Phys. Lett.}  B}

\def\l{\langle}
\def\r{\rangle}

\def\be{\begin{equation}}
\def\ee{\end{equation}}
\def\bea{\begin{eqnarray}}
\def\eea{\end{eqnarray}}

\newcommand{\Photo}{}

\begin{document}
\vspace*{4cm}
\title{IMPLICATIONS OF HIGGS DATA FOR\\[0.1cm] THE ELECTROWEAK CHIRAL LAGRANGIAN}

\author{ ALEJANDRO CELIS }

\address{ Ludwig-Maximilians-Universit\"at M\"unchen, 
   Fakult\"at f\"ur Physik,\\
   Arnold Sommerfeld Center for Theoretical Physics, 
   80333 M\"unchen, Germany }

\maketitle\abstracts{
In this talk I discuss the interpretation of LHC Higgs measurements within the electroweak chiral Lagrangian, paying particular attention to the role of power counting.      }

\section{Introduction}

Effective field theories (EFT) provide a powerful tool to study the low-energy dynamics of physical systems whenever a mass-gap is present.  In this case, the dynamics of the light degrees of freedom can be described in full generality without resorting to details of the ultraviolet (above the mass-gap) physics.   The low-energy effective theory consists of the most general Lagrangian respecting the underlying symmetries and involving the relevant light degrees of freedom.  Details of the ultraviolet dynamics are encoded in parameters of the effective theory that can be determined from experimental measurements at low-energies. This is particularly relevant since, in many situations, the ultraviolet dynamics is not known or its effects cannot be derived from first principles.  Such bottom-up approach has been proven extremely useful, for instance, in the study of pion interactions at low energies (below the $\rho$ mass).~\cite{Pich:1998xt}

In this talk I discuss the use of EFT methods for the interpretation of experimental measurements of the Higgs boson properties, with typical energies of the order of the electroweak scale $\sim 10^2$~GeV.   The presence of a mass-gap between the electroweak scale and the new dynamics, required for the formulation of an EFT, can be motivated based on the null results from direct and indirect searches for new physics performed so far.

In the electroweak sector, we can consider two bottom-up EFTs.  If the underlying dynamics responsible for electroweak symmetry breaking is weakly coupled, new-physics effects would decouple from the Standard Model and the expansion of the EFT is in canonical dimensions of the fields.~\cite{Weinberg:1979sa,Buchmuller:1985jz,Grzadkowski:2010es}    If the underlying dynamics is strongly coupled (expected around the TeV scale), new-physics effects do not decouple and the EFT can be organized as a loop expansion, or equivalently, according to the chiral dimension of fields and couplings.~\cite{ma3,ma7}   I will focus on the latter possibility.

\section{Electroweak Chiral Lagrangian}

The electroweak chiral Lagrangian describes an effective field theory, and as such, it is based on three pillars: {\textit{i)  relevant degrees of freedom,  ii) symmetries, and iii) an organizing principle or power counting.~\cite{ma3,ma7,Pich:2016lew}      }}  

\begin{description}
\item[i) Field content:]  Standard Model particle content.   It is assumed that the (transverse) gauge bosons and fermions are weakly coupled to the Higgs sector dynamics.   
\item[ii) Symmetries:]  Standard Model gauge symmetries plus conservation of lepton and baryon number; conservation of custodial symmetry at lowest order, CP invariance in the Higgs sector and fermion flavour. 

Three Goldstone bosons arise from the spontaneous breaking of the global symmetry $\mathrm{SU(2)}_L \otimes \mathrm{SU(2)}_R \to \mathrm{SU(2)}_V $. Since the factor $\mathrm{SU(2)}_L$ and the $\mathrm{U(1)}_Y$ subgroup of $\mathrm{SU(2)}_R$ are gauged, these Goldstone bosons are not present in the physical spectrum and instead become the longitudinal polarization components of the electroweak gauge bosons.     

Under a chiral transformation $\mathrm{SU(2)}_L \otimes \mathrm{SU(2)}_R$, the Goldstone boson matrix and the Higgs field transform as
\begin{equation}
U \to  g_L \, U \, g_R^{\dag} \,, \qquad h \to h \,, \qquad g_{L,R} \in \mathrm{SU(2)}_{L,R} \,.
\end{equation}
Here $U=\exp(2i\phi^a T^a/v)$, with $T^a$ representing the generators of $\mathrm{SU(2)}$ and $\phi^a$ the Goldstone fields.

\item[iii) Power Counting:]  The electroweak chiral Lagrangian is organized as an expansion in loops,~\cite{ma3,ma7} or equivalently in powers of $v^2/\Lambda^2$ where $v \simeq 246$~GeV represents the electroweak scale and $\Lambda \simeq 4 \pi v$ is the natural cut-off of the theory.~\cite{ma5} 

It is useful to introduce the concept of chiral dimension ($\chi$) for fields and couplings to keep track of the loop order in the electroweak chiral Lagrangian, just as in the study of low-energy pion dynamics.~\cite{ma4}  In this case we assign chiral dimension $0$ to bosons (gauge fields $X_{\mu}$, Goldstone fields $\varphi$, and the Higgs $h$) and $1$ for each derivative, weak coupling, and fermion bilinear.~\cite{ma3,ma7}  That is: $[ X_{\mu}, \varphi, h ]_{\chi} = 0$ and $[\partial_{\mu}, g, y, \psi \bar \psi]_{\chi} = 1$. 
Here $g$ and $y$ denote a generic gauge and Yukawa coupling, respectively.   The loop order $L$ of a given term in the effective Lagrangian is related to its chiral dimension through a simple relation: $\chi = 2 L +2$.
 
\end{description}

The electroweak chiral Lagrangian (EWChL) is organized as $\mathcal{L}_{\rm{EWChL}} =  \mathcal{L}_2 +    \mathcal{L}_4 + \cdots $,
where the subscript $n$ in $\mathcal{L}_n$ denotes the chiral dimension.    The {\textit{leading order}} (LO) of the electroweak effective theory has chiral dimension two and is given by~\cite{Contino:2010mh,Contino:2010rs,ma3} 
\begin{eqnarray}\label{eqlo}
{\cal L}_2 &=& -\frac{1}{2} \langle G_{\mu\nu}G^{\mu\nu}\rangle
-\frac{1}{2}\langle W_{\mu\nu}W^{\mu\nu}\rangle 
-\frac{1}{4} B_{\mu\nu}B^{\mu\nu}
+\bar q i\!\not\!\! Dq +\bar l i\!\not\!\! Dl
 +\bar u i\!\not\!\! Du +\bar d i\!\not\!\! Dd +\bar e i\!\not\!\! De 
\nonumber\\
&& +\frac{v^2}{4}\ \l D_\mu U^\dagger D^\mu U\r\, \left( 1+F_U(h)\right)
+\frac{1}{2} \partial_\mu h \partial^\mu h - V(h) \nonumber\\
&& - v \left[ \bar q \left( Y_u +
       \sum^\infty_{n=1} Y^{(n)}_u \left(\frac{h}{v}\right)^n \right) U P_+r 
+ \bar q \left( Y_d + 
     \sum^\infty_{n=1} Y^{(n)}_d \left(\frac{h}{v}\right)^n \right) U P_-r
  \right. \nonumber\\ 
&& \quad\quad\left. + \bar l \left( Y_e +
   \sum^\infty_{n=1} Y^{(n)}_e \left(\frac{h}{v}\right)^n \right) U P_-\eta 
+ {\rm h.c.}\right] \,.
\end{eqnarray}
The first line of \eqref{eqlo} represents the unbroken Standard Model while the following lines describe the sector of electroweak symmetry breaking.    We have denoted $P_{\pm} = 1/2 \pm T_3$.   The trace of a matrix $A$ is written as $\langle A \rangle$.    The left-handed doublets of quarks and leptons are denoted as $q$ and $l$, while the right-handed singlets are written as $u,d,e$.   In the Yukawa interactions, the right-handed fields are collected into $r = (u,d)^T$ and $\eta = (\nu, e)^T$.    The functions $F_U(h)$ and $V(h)$ take the form:
\begin{align}  \label{edjf}
F_U(h) = \sum_{n=1}^{\infty} f_{U,n}  \left(  \frac{h}{v} \right)^n \,, \qquad  V(h) =    v^4 \, \sum_{n=2}^{\infty} f_{V,n}  \left(  \frac{h}{v} \right)^n  \,.
\end{align}
The LO Lagrangian $\mathcal{L}_2$ is non-renormalizable.  The full theory however is renormalizable order by order in the chiral expansion.    One-loop divergences of the {\textit{leading order}} Lagrangian are absorbed by counter-terms present in the {\textit{next-to-leading order}} (NLO) piece $\mathcal{L}_4$.~\cite{ma6}    The NLO Lagrangian can be written generically as~\cite{ma3}
\begin{align}
   \mathcal{L}_4  = \sum_i c_i  \frac{v^{6-d_i}}{\Lambda^2}   \mathcal{O}_i     \,.
\end{align}
Here $d_i$ represents the canonical dimension of the operator $\mathcal{O}_i$.  The NLO operators come suppressed by $\Lambda^2 \simeq 1 6 \pi^2 v^2$ and have dimensionless coefficients $c_i$ which are naturally of order one.    Among the NLO operators,~\cite{ma3} those relevant for the single Higgs processes measured at the LHC are
\begin{align} \label{eqfg}
\mathcal{O}_{Xh1} &=  g^{\prime \, 2}  B_{\mu \nu}  B^{\mu \nu}  \,  F_{Xh 1}(h) \,,\qquad  \,~\,
\mathcal{O}_{Xh2} =  g^2 \langle W_{\mu \nu} W^{\mu \nu}  \rangle \, F_{Xh2}(h) \,, \nonumber \\
\mathcal{O}_{Xh3} &= g_s^2 \langle G_{\mu \nu} G^{\mu \nu} \rangle  \, F_{Xh3}(h) \,, \qquad 
\mathcal{O}_{XU1} =  g^{\prime} g B_{\mu \nu} \langle W^{\mu \nu} U T_3 U^{\dag} \rangle \, ( 1 + F_{XU1}(h)) \,,
\end{align}
with each of the functions $F_{Xhj}(h)$ and $F_{XU1}(h)$ having an analogous structure to that in Eq.~\eqref{edjf}.    Note that all the operators in Eq.~\eqref{eqfg} have chiral dimension four.

\section{Implications of Higgs Data}

The interpretation of Higgs data within the electroweak chiral EFT amounts to the determination of the EFT coefficients given current experimental data.  I discuss this problem using the methods of {\textit{Bayesian inference}}.   We are given an EFT described by an infinite number of operators, whose associated coefficients must be determined from the available measurements.  The EFT is predictive since it is organized via a power counting principle, such that at a given level of precision, only a finite subset of these operators is relevant.      This knowledge enters in the analysis in the form of Bayesian priors characterizing the expected size of the different contributions to a given process.     In our case, operators of the NLO Lagrangian enter at tree-level with a relative suppression of $v^2/\Lambda^2 \simeq 1/16 \pi^2$ compared to LO operators.~\cite{ma1,ma2}

\begin{figure}[ht]
\centering
\includegraphics[width=7cm]{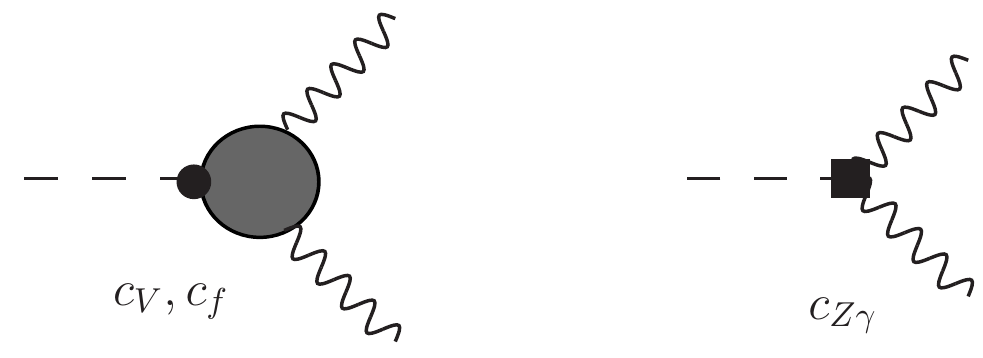} 
\caption{\small  \textit{Contributions to $h  \to Z \gamma$ involving an insertion from $\mathcal{L}_2$ (left) and the local term from $\mathcal{L}_4$ (right).   The gray blob in the left-diagram represents the different topologies of the SM-like loop diagrams. }      }
\label{fig:plot1c}
\end{figure}

I consider on-shell Higgs measurements performed with data from the first run of the LHC.   These are presented in the form of Higgs signal strengths and include Higgs decays into fermions $h \to \tau^+ \tau^-, b \bar b, \mu^+ \mu^-$ and electroweak gauge bosons $h \to \gamma \gamma, W^+W^-, ZZ, Z \gamma$.   The probed production mechanisms are gluon fusion, associated production with a vector boson, vector boson fusion, and associated production with top quarks.

Calculating the relevant processes up to the first non-trivial order in the electroweak chiral EFT amounts to considering the following terms~\cite{ma1,ma2} 
\begin{equation}\label{llopar2}
\begin{array}{ll}
  \mathcal{L} &=2 c_{V} \left(m_{W}^{2}W_{\mu}^{+}W^{-\mu} 
+\dfrac{1}{2} m^2_Z Z_{\mu}Z^{\mu}\right) \dfrac{h}{v} - \sum_f c_{f} y_{f} \bar{f} f h   \\
 &+ \dfrac{e g^{\prime}}{16\pi^{2}} c_{Z\gamma} Z_{\mu\nu}F^{\mu\nu} \dfrac{h}{v}  
 + \dfrac{e^{2}}{16\pi^{2}} c_{\gamma\gamma} F_{\mu\nu}F^{\mu\nu} \dfrac{h}{v}  
+\dfrac{g_{s}^{2}}{16\pi^{2}} c_{gg}\langle G_{\mu\nu}G^{\mu\nu}\rangle\dfrac{h}{v}  \,,
\end{array}
\end{equation}
where $y_f=m_f/v$.   Note that for loop-induced processes such as $gg \to h$ and $h \to \gamma \gamma, Z \gamma$, the loop contributions involving a vertex from $\mathcal{L}_2$ enter at the same order as the corresponding local contribution from $\mathcal{L}_4$ (see Fig.~\ref{fig:plot1c}).       At this level, the electroweak chiral EFT corresponds to the so called {\textit{kappa}} formalism.~\cite{ma1,kapa}  An extension of the analysis including higher order contributions can also be performed consistently with the power counting of the theory once the experimental precision requires it.~\cite{ma2}

\begin{figure}[ht]
\centering
\includegraphics[width=7cm]{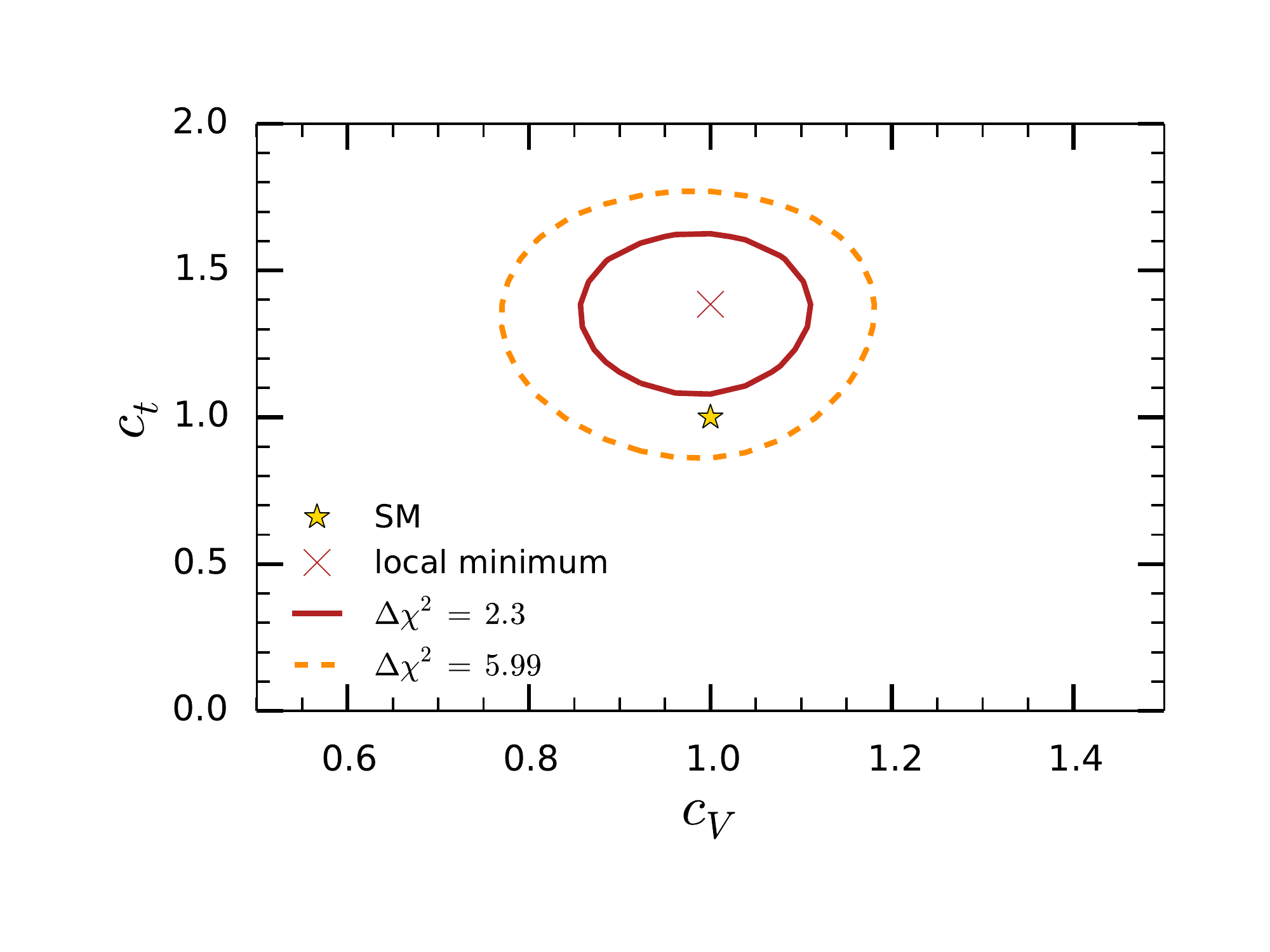} 
~
\includegraphics[width=7cm]{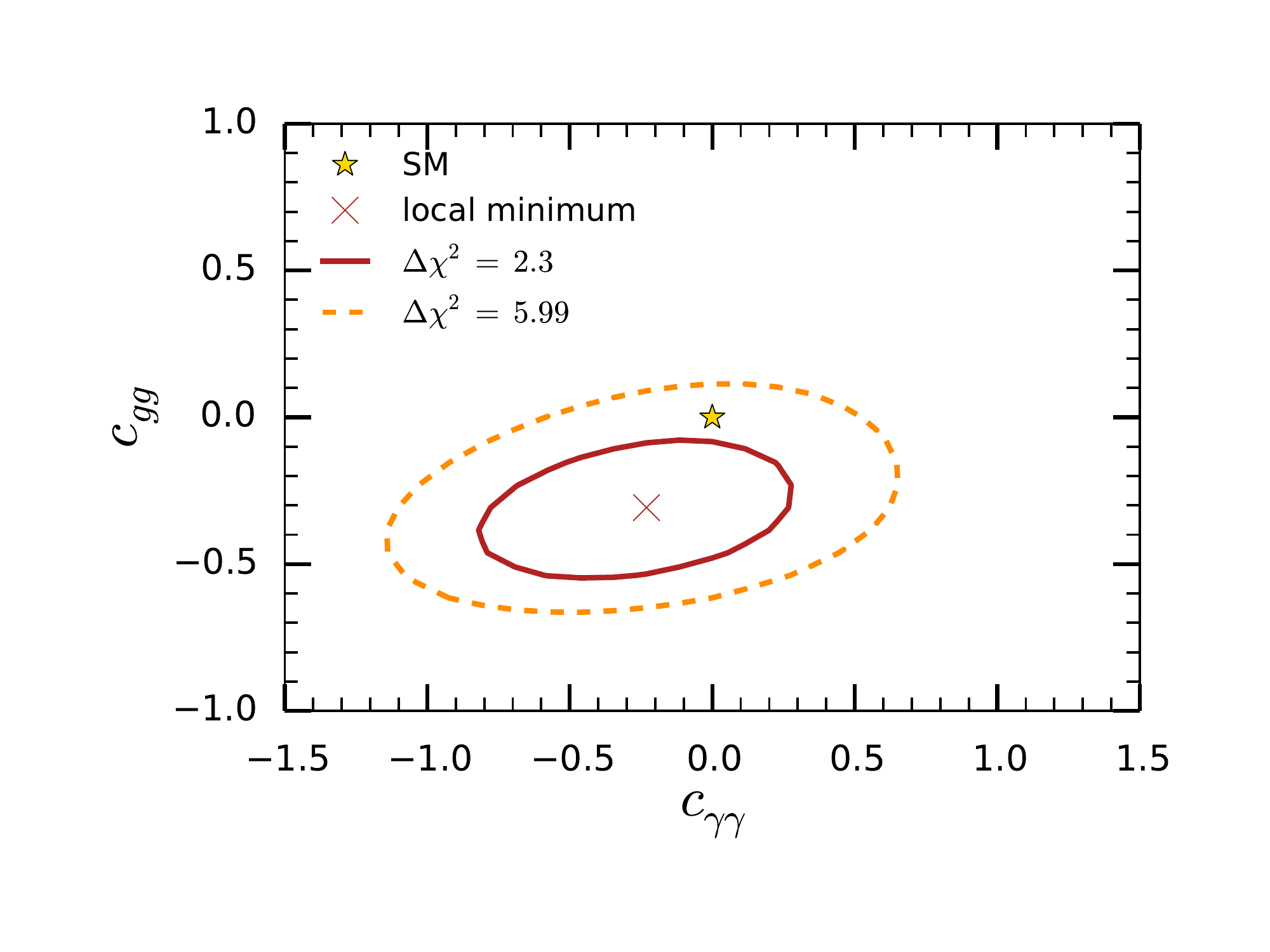} 
\vspace{-0.7cm}
\caption{\small  \textit{$68\%$ and $95\%$ Bayesian credible regions, approximated in this case by iso-contours of $\Delta \chi^2 = \chi^2 - \chi^2_{\rm{min}}$ with $\chi^2 \equiv -2 \mathrm{log} \text{(pdf)}$.      }}
\label{fig:plot2}
\end{figure}

The coefficients $c_i$ in \eqref{llopar2} are naturally of order one.    This knowledge can be taken into account by defining ``appropriate" prior probability density functions (pdf) which are then folded with the Likelihood in order to obtain the posterior pdf.       Fig.~\ref{fig:plot2} shows $68\%$ and $95\%$ Bayesian credible regions in the planes $c_V$-$c_t$ and $c_{\gamma \gamma}$-$c_{gg}$ for illustration, taking flat priors in the ranges: $c_V \in [0.5,1.5]$, $c_{f} \in [0,2]$, $c_{\gamma \gamma, Z \gamma} \in [-1.5, 1.5]$, $c_{gg} \in [-1,1]$.

\section*{Acknowledgments}

I am grateful to the organizers of the Rencontres de Blois for the pleasant atmosphere.    I thank G.~Buchalla, O.~Cat\`a and C.~Krause for the fruitful collaboration. 
I acknowledge support from the Alexander von Humboldt Foundation.

\end{document}